\documentclass[a4paper,12pt]{article}
\usepackage{graphicx}
\usepackage{amssymb}
\usepackage{citesort}
\usepackage{latexsym}
\usepackage{psfrag}
\newsavebox{\LSIM}
\sbox{\LSIM}{\raisebox{-1ex}{$\ \stackrel{\textstyle<}{\sim}\ $}}

\newsavebox{\GSIM}
\sbox{\GSIM}{\raisebox{-1ex}{$\ \stackrel{\textstyle>}{\sim}\ $}}

\begin{document}
\begin{titlepage}
\begin{flushright}
\end{flushright}
$\mbox{ }$
\vspace{.1cm}
\begin{center}
\vspace{.5cm}
{\bf\Large The bubble wall velocity in the minimal supersymmetric light stop scenario }\\[.3cm]
\vspace{1cm}
Stephan J.~Huber$^{a,}$\footnote{s.huber@sussex.ac.uk}
and 
Miguel Sopena$^{a,}$\footnote{m.sopena@sussex.ac.uk} \\ 
\vspace{1cm} {\em  
$^a$ Department of Physics and Astronomy, Sussex University, Brighton, East Sussex BN1 9QE, UK}\\[.2cm] 
\end{center}
\vspace{1.cm}
\begin{abstract}
We build on existing calculations of the wall velocity of the expanding bubbles of the broken symmetry phase in a first-order electroweak phase transition within the light stop scenario (LSS) of the MSSM. We carry out the analysis using the 2-loop thermal potential for values of the Higgs mass consistent with present experimental bounds. 
Our approach relies on describing the interaction between the bubble and the hot plasma by a single friction parameter, which we fix by matching to an existing 1-loop computation and  extrapolate it to our regime of interest.
For a sufficiently strong phase transition (in which washout of the newly-created baryon asymmetry is prevented) we obtain values of the wall velocity, $v_w\approx0.05$, far below the speed of sound in the medium, and not very much deviating from the previous 1-loop calculation. We also find that the phase transition is about 10\% stronger than suggested by simply evaluating the thermal potential at the critical temperature.
\end{abstract}
\end{titlepage}
\section{Introduction}
It has long been argued that Sakharov's three conditions for baryogenesis \cite{Sakharov:1967dj} (deviation from thermal equilibrium, CP violation, and baryon number violation) could have been met in the context of a first-order electroweak phase transition (EWPT) at the $\sim100$ GeV scale \cite{Kuzmin:1985mm} (for a review, see e.g.~\cite{Cline:2002aa}). Baryon number violation in this setting would have proceeded through so-called 'sphaleron' transitions, which are Boltzmann-suppressed and become slow after the EWPT at $T \lesssim 100$ GeV, with the sphaleron energy $E_{\rm Sph}/T \sim v/T$, $v$ being the Higgs VEV. In a first-order transition separate regions of the old and the new phase coexist, with bubbles of the new phase nucleating and expanding in the old phase until they fill all space. Baryon number violation would have taken place outside the growing bubbles, and then the baryon asymmetry would have been transported across the bubble wall into the new phase, where it must avoid annihilation by washout processes. This is granted by a sufficiently strong phase transition with $v/T\gtrsim1.0$ \cite{Moore:1998swa}, which is impossible to satisfy within the Standard Model for experimentally allowed values of the Higgs mass \cite{Kajantie:1996mn}. Therefore extensions of the Standard Model, such as models with extra Higgs fields \cite{Fromme:2006cm,Cline:2011mm} or singlets \cite{Pietroni:1992in,Huber:2000mg,Menon:2004wv,Huber:2006ma,Espinosa:2011ax,Espinosa:2011eu}, or non-standard Higgs potentials \cite{Grojean:2004xa,Bodeker:2004ws} are required for successful electroweak baryogenesis.

The possibility of a strong electroweak phase transition in a minimal supersymmetric setting leading to the production of a baryon asymmetry consistent with observations has been studied extensively. Electroweak baryogenesis has been shown to be feasible in a specific region in the supersymmetric mass parameter space. This setting is generally known as the light stop scenario (LSS), characterised by a (predominantly right-handed) light stop with a mass lighter than or comparable to that of the top quark \cite{Carena:1996wj,Bodeker:1996pc,Laine:1998qk,Carena:2008vj}. All other squarks and sleptons are typically taken to be at a  much higher mass scale. Being one more light bosonic species (in addition to the weak gauge bosons) that couples to the Higgs, the light stop increases the upper bound on the Higgs mass compatible with a strong phase transition to about 127 GeV \cite{Carena:2008vj}. In turn, the predominantly left-handed stop must be heavy to agree with electroweak precision tests and to provide a sufficiently heavy Higgs boson. Gluinos are generally considered heavy and thus decoupled from the thermal bath in order to suppress their potentially large contribution to the effective thermal light stop mass. Charginos and neutralinos should remain light as they provide the only additional CP-violating currents available in this context, needed to generate the observed baryon asymmetry of the Universe \cite{Huet:1995sh,Carena:2002ss,Konstandin:2005cd,Chung:2008aya}. The CP-odd Higgs mass is large to avoid potentially large contributions to the electric dipole moments of the electron and the neutron, leaving one light, SM-like Higgs boson.

A crucial parameter entering the computation of the generated baryon asymmetry \cite{Huet:1995sh,Carena:2002ss,Konstandin:2005cd,Chung:2008aya} is the velocity of the expanding bubble walls, $v_w$. In particular, it is essential that the wall velocity is smaller than the speed of sound in the plasma, so that diffusion of charges from the bubble wall into the symmetric phase is possible (see however ref.~\cite{Caprini:2011uz} for a recent proposal for supersonic electroweak baryogenesis).

The wall velocity is determined by the friction induced by the motion of the bubble through the plasma, and by the pressure difference across the bubble wall. Microscopically, friction is related to deviations from equilibrium in the plasma. In a semiclassical approximation this effect can be described by a set of Boltzmann equations coupled to the equations of motion of the Higgs field. Such a system of equations was written down and solved for the first time in ref.~\cite{Moore:1995si} in the context of small Higgs masses in the minimal Standard Model. Later on the computation was generalized to the LSS in ref.~\cite{John:2000zq}. While the typical wall velocity in the Standard Model was found to be $v_w\sim0.35$, its value in the LSS was determined to be almost an order of magnitude smaller. This is due to additional friction induced by the light stops.

Ref.~\cite{John:2000zq} makes use of the thermal Higgs potential in 1-loop approximation, while 2-loop corrections are known to be vital to obtain a strong phase transtion for realistic Higgs masses \cite{Carena:1996wj,Bodeker:1996pc,Laine:1998qk,Carena:2008vj}. The aim of the present work is to extend the computation of the wall velocity in the LSS to the case of the 2-loop thermal potential, resulting in a sufficiently strong phase transition for experimentally allowed Higgs masses around and above 115 GeV.


Rather than repeating the microscopic analysis of friction from the literature, we model friction by a single friction parameter, which we fix by matching to ref.~\cite{John:2000zq} and extrapolate it to our Higgs mass range of interest. As we will show such an approach leads to a much simpler set of equations, basically relativistic hydrodynamics coupled to the equation of motion of the Higgs field \cite{Ignatius:1993qn}. This method has been successfully used in ref.~\cite{KurkiSuonio:1996rk} to numerically simulate bubble growth. The main idea of our work is that friction is mainly related to interactions between the wall and the hot plasma, and does therefore hardly change when higher-order corrections are included in the thermal potential. On the other hand, friction would indeed change significantly, if the composition of the plasma would change, e.g. by removing the light stops.


We will show that the main result of  ref.~\cite{John:2000zq} carries over to the 2-loop case: bubble walls in the LSS move very slowly, $v_w\approx0.05$, far below the speed of sound in the medium. This is also true for Higgs masses around and above 115 GeV, and for a phase transition strong enough to avoid baryon number washout. As a by-product we also arrive at a more reliable determination of the strength of the phase transition, reading off the Higgs VEV inside a realistic bubble, including effects of reheating, rather than taking it from the minimum of the thermal potential at the critical temperature. Thus we find that the relevant value of $v/T$ is increased by about 10\%.


\section{Calculation of the wall velocity}

\subsection{The perfect fluid setting and the hydrodynamic equations}

We treat the problem of the moving bubble wall by modelling the hot plasma as a perfect relativistic fluid, assuming conservation of its energy-momentum tensor, which is the sum of the separate contributions from the fluid and the  Higgs field \cite{Ignatius:1993qn}:
\begin{eqnarray}
\partial_{\mu}T^{\mu\nu}=\partial_{\mu}(T_{\rm field}^{\mu\nu}+T_{\rm fluid}^{\mu\nu})=\nonumber \\
=\partial_{\mu}\Big(\partial^{\mu}\varphi\partial^{\nu}\varphi-g^{\mu\nu}(\frac{1}{2}\partial_{\alpha}\varphi\partial^{\alpha}\varphi)+\nonumber \\
+(\rho+P)u^{\mu}u^{\nu}-Pg^{\mu\nu}\Big)=0.
\end{eqnarray}
Here  $\rho$ the energy density, $P$ the pressure, and $g^{\mu\nu}$ the usual Minkowski metric. 

Stationary\footnote{Steady-state solutions to the hydrodynamic equations for a spherical bubble are \textit{similarity solutions}, that is, they maintain their relative shape but rescale as the bubble grows.} solutions to the problem of the expanding bubble wall are usually divided into two categories depending on whether the wall advances at a velocity above or below the speed of sound in the medium \cite{Steinhardt:1981ct,Laine:1993ey,Espinosa:2010hh}. In subsonic solutions ('deflagrations') the bubble wall is preceded by a 'shock front' that accelerates and heats up the plasma, which is brought to rest by the bubble wall passing through. In supersonic solutions ('detonations') the plasma is hit by the bubble wall while at rest and accelerated, and is brought back to rest by a rarefaction wave which follows the wall. For the purposes of this work velocities remain well below the speed of sound and we will have no cause to explore supersonic solutions to the hydrodynamic equations.

It is convenient to work with the stress-energy conservation equation in the rest frame of the expanding bubble wall, picking the radial direction and approximating the situation at the bubble wall as planar. We are left with one spatial coordinate and no time dependence. Applying the relevant thermodynamic relations and introducing our friction coefficient $\eta$ we arrive at the coupled system  \cite{Ignatius:1993qn,Sopena:2010zz}
\begin{eqnarray}
\frac{d^2\varphi(x)}{dx^2}=\frac{\partial V(\varphi,T)}{\partial\varphi}+\eta\frac{\varphi^2}{T_{s1}}v\gamma\frac{d\varphi(x)}{dx}\label{eqn:fluid1}\\
(4aT^4-T\frac{\partial V(\varphi,T)}{\partial T})\gamma^2v=C_1\label{eqn:fluid2}\\
(4aT^4-T\frac{\partial V(\varphi,T)}{\partial T})\gamma^{2}v^{2}+P_{r}-V(\varphi,T)+\frac{1}{2}(\frac{d\varphi}{dx})^{2}=C_2\label{eqn:fluid3}
\end{eqnarray}
where $\varphi$
 is the scalar field, $v$  is the fluid velocity, $T$ the temperature, $\gamma$ the relativistic factor $(1-v^2)^{-1/2}$, $T_{s1}$ the plasma temperature in the symmetric phase ahead of the advancing wall, and $C_1$, $C_2$ are integration constants best determined in the symmetric phase where most contributions vanish. The radiative pressure is $P_{r}= aT^{4}=\frac{\pi^{2}}{90}g^{*}T^{4}$, $g^{*}$ being the number of effective degrees of freedom in the plasma at the temperature $T$. The special form of the friction term in (\ref{eqn:fluid1}) is motivated by the microscopic analysis of ref.~\cite{Moore:1995si}.

This system can be solved numerically across the bubble wall by imposing vanishing $\varphi$ derivatives at both boundaries of the integration interval and vanishing Higgs VEV in the symmetric phase. We continue the computation of the bubble profile by solving the hydrodynamic equation in the region between the bubble wall and the shock front, for which we drop the planar approximation and take into account the sphericity of the bubble (note that on this step we are already in the symmetric phase where we take $\varphi=V(\varphi, T)=0$). Lastly we solve the conservation equation across the shock front to relate our findings to the temperature of the universe unperturbed by the bubble \cite{Espinosa:2010hh}.

As usual, a first-order phase transition is indicated by the effective potential at the relevant temperature, which has a global minimum at zero Higgs VEV, developing a second local minimum at nonzero VEV as the temperature of the universe decreases (see e.g.~\cite{Anderson:1991zb}). As $T$ keeps decreasing the value of $V$ at the second minimum approaches the value of the symmetric minimum. The minima become degenerate at the so-called critical temperature, $T_c$. Nucleation of bubbles of the new phase becomes possible for $T<T_c$. The phase transition is deemed to begin once the integrated probability of bubble nucleation in the horizon volume 
\begin{eqnarray}
P(T)=\int^{T_c}_{T}dP=\int^{T_c}_{T}\left(\Gamma/\textit{Vol}\right)\cdot{V}_H\cdot{dt}=\int^{T_c}_{T}\frac{T^4}{H^4}e^{-F_{c}/T}\frac{dT}{T}\label{eqn:nuctemp}
\end{eqnarray}
reaches unity, at the nucleation temperature $T_n$ (here we have taken $\Gamma/\textit{Vol}=\Lambda^{4}(T)e^{-F_{c}/T}\approx T^4 e^{-F_{c}/T}$, $F_c$ being the free energy of the so-called critical bubble, just large enough to spontaneously grow without collapsing under surface tension, at the temperature $T$).
Once bubbles begin to nucleate they grow and occupy the whole space very quickly, ending the phase transition.

\subsection{The wall velocity in the MSSM}

The wall velocity of expanding bubbles of the broken symmetry phase in a 1st-order electroweak phase transition was calculated microscopically for the Standard Model by Moore and Prokopec  \cite{Moore:1995si} and for the MSSM, based on the same procedure, by John and Schmidt \cite{John:2000zq}. The equation of motion for one scalar background field coupled to the distribution functions of the particles in the plasma is \cite{Moore:1995si}
\begin{equation} \label{eomMP}
\Box \varphi+\frac{\partial V(\varphi)}{\partial \varphi}+\Sigma\frac{dm^2}{d\varphi}\int\frac{d^3p}{(2\pi)^32E}f(p,x)=0.
\end{equation}
Here $V(\varphi)$ is the renormalised vacuum potential, the sum is over all particle species, and the mass dependence of each particle on the Higgs VEV is given by its respective couplings. The vacuum potential combines with the equilibrium part of the particle distribution to produce the finite-temperature effective potential $V_{eff}(\varphi,T)$. The equation of motion can then be written as
\begin{equation} \label{eomMP2}
\Box \varphi+\frac{\partial V_{\rm eff}(\varphi,T)}{\partial \varphi}+\Sigma\frac{dm^2}{d\varphi}\int\frac{d^3p}{(2\pi)^32E}\delta f(p,x)=0
\end{equation}
where $\delta f$ expresses the deviation from equilibrium of the relevant particle populations, which is responsible for the wall friction. The evolution of the particle populations, in the semiclassical approximation, is given by Boltzmann equations of the form
\begin{equation}
\frac{df}{dt}=\partial_tf+\dot{\vec{x}}\cdot\partial_{\vec{x}}f+\dot{\vec{p}}\cdot\partial_{\vec{p}}f=-C[f];
\end{equation}
$C[f]$ being the \textit{collision integral}. These Boltzmann equations together with the equation of motion for the scalar field constitute a very complicated set of coupled integro-differential equations, which is very difficult to solve.

Several simplifications are possible in order to solve the equation of motion. Taking $C[f]\equiv0$ constitutes the 'free particle' approximation. In this approach the Boltzmann equation can be solved exactly  \cite{Moore:1995si} and $f$ integrated over momentum. Recently this approximation was used in ref.~\cite{Bodeker:2009qy} to show that under certain circumstances the bubble wall velocity may approach the speed of light.
In the 'relaxation time approximation' $C[f]\equiv(\frac{\delta f}{\tau})$, with $\tau$ independent of momentum. This approach becomes particularly simple if $\tau>>L_w$ with $L_w$ the wall thickness, because then the spatial derivatives of $\delta$ can be neglected.

The approach first followed by Moore and Prokopec for the Standard Model \cite{Moore:1995si}, then by John and Schmidt for the MSSM \cite{John:2000zq} (the 'fluid approximation') describes $f$ for each relevant particle as an equilibrium distribution where $E/T$ is modified by perturbations,
\begin{equation}
f_i \equiv f_0(E+\delta) = \frac{1}{\exp(\frac{E+\delta}{T})\pm1},
\end{equation}
where $\delta=-[\delta\mu + \delta\mu_{bg} + \frac{E}{T}(\delta T + \delta T_{bg}) + p_z(\delta v + \delta v_{bg})]$.
It is sufficient to treat in this way only 'heavy' species which couple strongly to the Higgs (top quarks, W bosons, and Z bosons for the Standard Model, and additionally right-handed stops in the MSSM). Note that in the MSSM, as mentioned, other ('superheavy') particles are taken as decoupled from the plasma and show up only in the shape of the effective vacuum potential of the theory. The treatment of 'light' particles is simplified by treating them as one species with common perturbations to the equilibrium chemical potential, temperature and fluid velocity $\delta\mu_{bg}$, $\delta T_{bg}$ and $\delta v_{bg}$ (as opposed to the individual perturbations for 'heavy' species $\delta\mu$, $\delta T$ and $\delta v$).

With this approximation the Boltzmann equation for each particle species can be expanded to linear order in the perturbations and then $\int d^3p / (2\pi)^3$, $\int E d^3p / (2\pi)^3$, and $\int p_z d^3p / (2\pi)^3$ integrations carried out. This, plus the fluid equation, constitutes the system to be solved numerically \footnote{John and Schmidt, in the MSSM case, refer explicitly to the simplification in which all $\delta'=0$. In that case a fluid equation for each scalar field formally identical to (\ref{eqn:fluid1}) is produced with a calculable friction parameter $\eta$.}.


\section{The MSSM 2-loop potential}

As the model's effective theory we base ourselves on the 2-loop, finite temperature MSSM potential calculated in refs.~ \cite{Carena:1996wj,Bodeker:1996pc,Laine:1998qk,Carena:2008vj,Espinosa:1996qw} with only one (light) background field with SM-like couplings to vector bosons and fermions. Only third generation squarks are considered to be at the electroweak scale, and we assume no mixing between right- and left-handed stops. With this the additional supersymmetric parameters of the model are just the soft supersymmetry-breaking mass parameters for the left- and right-handed stop $m_Q$ and $m_U$ and $\tan\beta=\frac{v_2}{v_1}$ (with $v_1$, $v_2$ being the zero temperature expectation values of the real parts $\phi_1$, $\phi_2$ of the neutral components of the two supersymmetric Higgs doublets). In the present case of only one light Higgs field the effective potential can be expressed, as in the Standard Model, as a function of only one background field.

\subsection{The 1-loop potential}

The 1-loop portion of the effective potential for the light Higgs field $\varphi$ is \cite{Espinosa:1996qw}
\begin{eqnarray}
V_{\rm tree}(\varphi) = - \frac{1}{2} \mu^2 \varphi^2 + \frac{1}{32} \varphi^4 g^2 \cos^2 2\beta\\
V_{\rm 1-loop, 0-T}(\varphi) = \sum \frac{n_i}{64 \pi^2}m_i^4(\varphi) [\log \frac{m_i^2(\varphi)}{Q^2}-C_i]\\
V_{\rm 1-loop, thermal}(\varphi,T) = \frac{T^4}{2 \pi^2} \sum n_i J_i[\frac{\bar{m}_i^2(\varphi,T)}{T^2}]
\end{eqnarray}
Note that we take $g'=0$. Since 1-loop contributions to the potential are comparable to the tree-level portion, the parameter $\mu^2$ in the tree-level part is chosen so that the minimum of the total 1-loop, non-thermal potential lies at $\varphi_0=245.7$ GeV.

Sums run over all species that contribute significantly. For the 1-loop part this includes stops, tops and W and Z bosons. The left-handed stops do not contribute to the  thermal piece. The number of degrees of freedom for each species, $n_i$, is
\begin{equation}
n_t=-12,\;n_{\tilde{t}_R}=n_{\tilde{t}_L}=6,\;n_W=6,\;n_Z=3
\end{equation}
We take $Q=m_Z$ and $C_i=\frac{5}{6}$ for vector bosons, $\frac{3}{2}$ for scalars and fermions. The relevant expansion of the $J$ functions in the 1-loop, thermal bit is of the form
\begin{eqnarray}
J_{\rm bosons}\left(\frac{\bar{m}^2_i}{T^2}\right)\equiv\int^{\infty}_{0} dx\;x^2 \log \left(1 - \exp\left(-\sqrt{x^2+(\frac{\bar{m}^2_i}{T^2})}\right)\right)=\nonumber\\
 = ( 2 \pi^2 ) \left( \frac{2}{48} \left(\frac{\bar{m}^2_i}{T^2}\right)^2 - \frac{1}{12 \pi} \left(\frac{\bar{m}^2_i}{T^2}\right)^3 - \frac{1}{64 \pi^2} \left(\frac{\bar{m}^2_i}{T^2}\right)^4 \log\left(\left(\frac{\bar{m}^2_i}{T^2}\right)^2 - 5.408\right) \right)\\
J_{\rm fermions}\left(\frac{m^2_i}{T^2}\right)\equiv\int^{\infty}_{0} dx\;x^2 \log \left(1 + \exp\left(-\sqrt{x^2+(\frac{m^2_i}{T^2})}\right)\right)=\nonumber\\
 = - ( 2 \pi^2 ) \left( \frac{1}{48} \left(\frac{m^2_i}{T^2}\right)^2 + \frac{1}{64 \pi^2} \left(\frac{m^2_i}{T^2}\right)^4 \log\left(\left(\frac{m^2_i}{T^2}\right)^2 - 2.635\right) \right)
\end{eqnarray}
The notation $\bar{m}_i$ (for bosons) in the 1-loop, thermal piece indicates resummed masses. $\bar{m}_i^2$ are obtained from $m^2_i$ by adding the leading temperature-dependent self-energy contributions (see below). An effect of the 2-loop calculation is to resum all masses in the thermal piece of the 1-loop potential\footnote{In the 1-loop version of our potential (which we use to reproduce the results of ref.~\cite{John:2000zq}) the masses in the 1-loop, T-dependent piece are not resummed. Following ref.~\cite{Espinosa:1996qw} we do resum the bosonic masses (only the longitudinal degrees of freedom for the gauge bosons, photons included) in the term cubic in $m$ in the expansion, through the addition of the piece $\Delta V (\varphi,T) = -\frac{T}{12 \pi} \Sigma n_i [\bar{m}_i^3(\phi,T) - m_i^3(\varphi,T)]$ running over the relevant species with $n_{W_L}=2$, $n_{Z_L}=n_{\gamma_L}=1$.}.

The expressions for the particle masses are:
\begin{eqnarray}
m^2_{\rm top}(\varphi)=\frac{1}{2} h^2_t \varphi^2 \sin^2 \beta,\\
m^2_{\tilde{t}_L}(\varphi) = m_Q^2 + m^2_t(\varphi) + D^2_{\tilde{t}_L}(\varphi),\\
m^2_{\tilde{t}_R}(\varphi) = m_U^2 + m^2_t(\varphi) + D^2_{\tilde{t}_R}(\varphi)
\end{eqnarray}
assuming no mixing between left- and right-handed stops and taking
\begin{eqnarray}
D^2_{\tilde{t}_L}(\varphi) = \frac{1}{4} \left(\frac{1}{2}-\frac{2}{3} \sin^2 \theta_W\right) g^2 \varphi^2 \cos 2\beta,\\
D^2_{\tilde{t}_R}(\varphi) = \frac{1}{4} \left(\frac{2}{3} \sin^2 \theta_W\right) g^2 \varphi^2 \cos 2\beta
\end{eqnarray}
Note that with the choice $g'=0$ we have $D^2_{\tilde{t}_R}\equiv 0$ for all $\varphi$.

As field-dependent mass for the W and Z gause bosons we take
\begin{equation}
m_W^2(\varphi) = m_Z^2(\varphi) = \frac{1}{4} g^2 \varphi^2.
\end{equation}

\subsection{The 2-loop contributions}

Following the notation of ref.~\cite{Espinosa:1996qw}, the relevant Standard Model 2-loop contribution to the potential can be written as
{\small
\begin{eqnarray}
V^{(2)}_{SM}&=&\frac{g^2}{16 \pi^2} T^2 \left[M^2\left(\frac{3}{4}\log\frac{M_L}{T}-\frac{51}{8}\log\frac{M}{T}\right)+\frac{3}{2}\left(M^2-4M_L^2\right)\log\frac{M+2M_L}{3T}+3MM_L\right]\nonumber\\
&&+\frac{m_t^2(\varphi)T^2}{64\pi^2}\left[16g_s^2\left(\frac{8}{3}\log2-\frac{1}{2}-c_B\right)+9h_t^2 \sin^2\beta\left(\frac{4}{3}\log2-c_B\right)\right]
\end{eqnarray}}
where $c_B=\log(4\pi)-\gamma_E$, $\gamma_E\approx0.577215665$ being Euler's constant. We take the strong coupling and top Yukawa coupling respectively as $g_S\approx1.228$ and $h_t\equiv\frac{\sqrt{2} m_{\rm top}}{\varphi_0 \sin\beta}$.
Here $M^2=\frac{1}{4} g^2 \varphi^2$ is the weak gauge boson mass and $M_L^2 = M^2+\frac{7}{3}g^2T^2$ the longitudinal resummed mass (corrected by the self-energy).

We have retained the supersymmetric contributions which relate to non-decoupled species in the plasma. The relevant diagrams are shown in \cite{Espinosa:1996qw}. We take as the total supersymmetric, 2-loop part of the potential for our calculation
\begin{eqnarray}
V_{\rm 2-loop,\;MSSM} = -\frac{g_s^2 (N_c^2-1)\;T^2}{16 \pi^2}\left(\bar{m}^2_{\tilde{t}_R}\log\frac{2\bar{m}_{\tilde{t}_R}}{3T}\right)+\nonumber\\
+ \frac{N_c \varphi^2 T^2}{32 \pi^2}\left(h_t^2\sin^2\beta\right)^2\log\frac{\bar{m}_h+2\bar{m}_{\tilde{t}_R}}{3T}-\nonumber\\
-\frac{g_s^2T^2}{64\pi^2}(N_c^2-1)(c_2-1) \bar{m}^2_{\tilde{t}_R}+\nonumber\\
+\frac{3N_c}{128\pi^2}T^2 h_t^4 \sin^4\beta c_2 \phi^2+\nonumber\\
+\frac{N_c T^2}{16\pi^2}\left[\frac{g_S^2}{6}(N_c+1)\bar{m}^2_{\tilde{t}_R}+\frac{1}{2}h_t^2\sin^2\beta\left(\bar{m}_h\bar{m}_{\tilde{t}_R}+3\bar{m}_{\chi}\bar{m}_{\tilde{t}_R}\right)\right]
\end{eqnarray}
We take the number of colours $N_c$ as $3$, and $c_2\approx3.3025$. 
Note that, given the large 1-loop corrections to the Higgs mass, $\bar{m}_h$ and $\bar{m}_\chi$ are expressed on the basis on taking $m_h=\frac{\partial^2V_{\rm 1-loop}}{\partial\varphi^2}$ at the minimum of the 1-loop potential.

\begin{figure}
\begin{center}
\psfrag{xi = 0.7}{$\xi = 0.7$}
\psfrag{xi = 0.8}{$\xi = 0.8$}
\psfrag{xi = 0.9}{$\xi = 0.9$}
\psfrag{xi = 1.0}{$\xi = 1.0$}
\psfrag{xi = 1.2}{$\xi = 1.2$}
\includegraphics{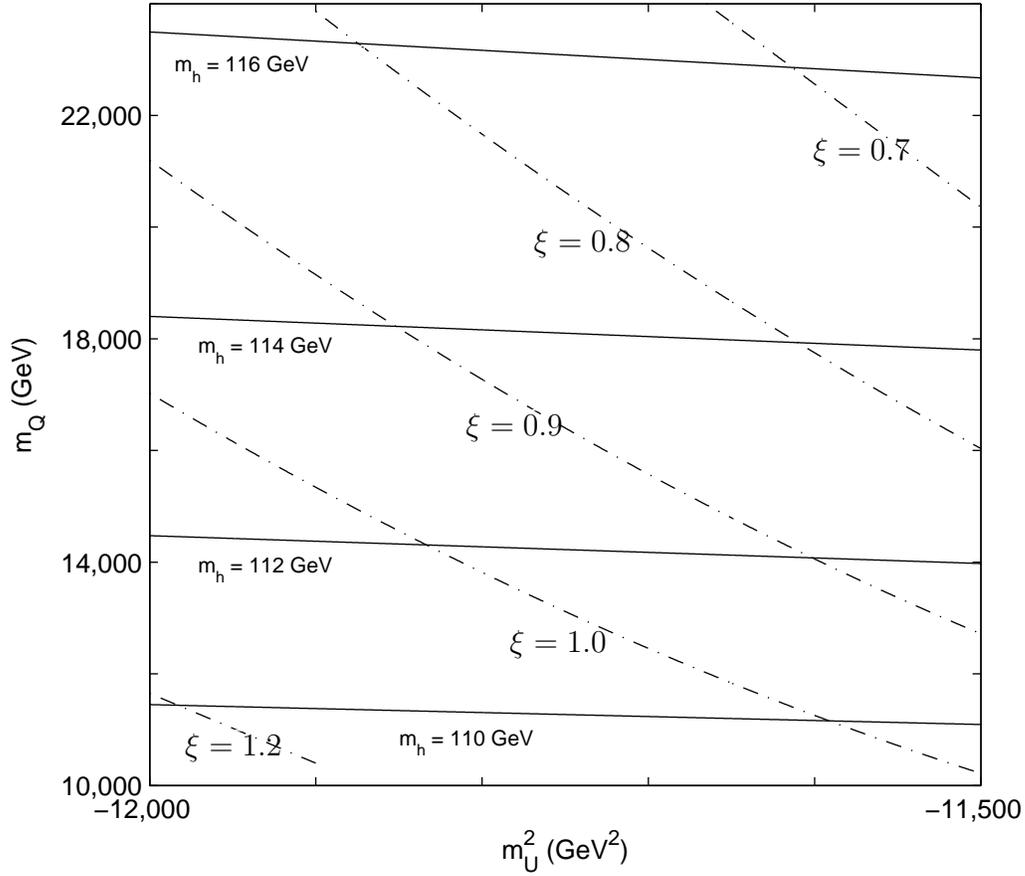}
\caption{Values of the strength of the phase transition $\xi = \frac{v_c}{T_c}$ and the Higgs mass in the region of parameter space (with $\tan \beta = 4$) of interest for baryogenesis. For this value of $\tan \beta$ the right-handed stop mass (we assume no mixing) varies in this range from $m_{\tilde{t}_R} \sim 135.2$ to $m_{\tilde{t}_R} \sim 138.8$ GeV and the left-handed-stop mass is given by $m_Q$ in units if GeV.}
\label{fig:MSSMparamspace}
\end{center}
\end{figure}

The region in $m_U^2$-$m_Q$ space for $\tan \beta = 4$ which provides both a sufficiently strong phase transition and presently acceptable values of the Higgs mass is shown in figure \ref{fig:MSSMparamspace}. We observe that a strong phase transition, i.e.~$\xi = \frac{v_c}{T_c}>1$, can be achieved for a light stop mass around 135 GeV and have stop mass above 10 TeV. We will see later on that this procedure underestimates the strength of the phase transition by about 10\%. Here the Higgs mass is around 115 GeV. Somewhat larger Higgs masses can be obtained by using larger values of $\tan \beta$ and/or $m_Q$. 

\section{Calibration of the friction parameter}

\begin{figure}
\begin{center}
\includegraphics{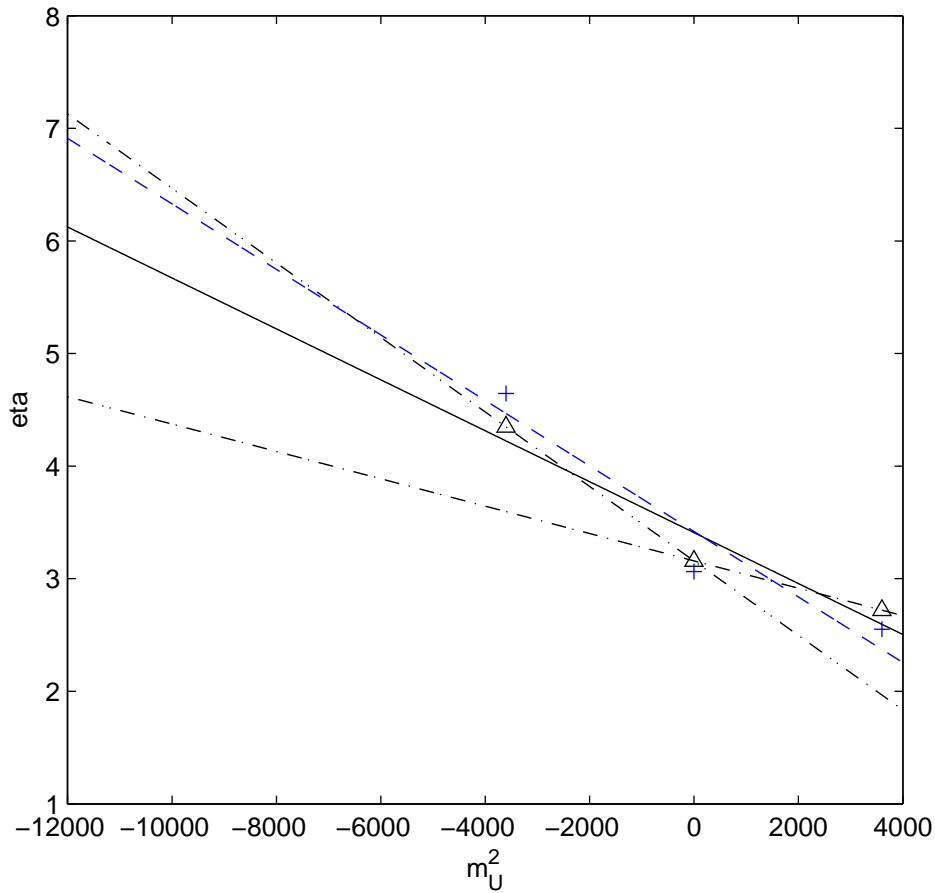}
\caption{Linear extrapolation of $\eta$ values to the $m_U^2$ region of interest for $\tan\beta=4$ (solid line, calibration values from \cite{John:2000zq} as triangles), $\tan \beta = 6$ (dashed line, calibration values as crosses). $m_{\tilde{t}_R} \in [135.2, 184.1]$ GeV (almost no dependence on $\tan\beta$)}.
\label{fig:extretatgbeta46}
\end{center}
\end{figure}

Our goal is to calculate the wall velocity in the region of parameter space which provides a sufficiently strong phase transition and is compatible with present bounds on the mass of a SM-like Higgs boson. We assume that the phenomenological friction parameter $\eta$ depends exclusively on $m_U$ and $\tan \beta$, as these are the parameters which govern the abundance and coupling of the light stops. We match the 1-loop results of John and Schmidt through the system (\ref{eqn:fluid1})-(\ref{eqn:fluid3}) to determine our friction parameter $\eta$ for $\tan \beta\in[2,6]$, $m_U\in[-60,+60]$ GeV. The coupled system can be linearised and solved numerically. The fitted $m_U$, $\tan \beta$-dependent friction parameter can then be extrapolated to the relevant region in MSSM parameter space. The values of the fitted friction parameter which reproduce the results of John and Schmidt are shown in table \ref{table:fitted etas}. We find that $\eta$ is almost independent of $\tan\beta$, but grows significantly with lower stop masses, i.e.~lower $m_U^2$. The linear extrapolation to the $m_U$ interval of interest for $\tan \beta=4$,$6$ is shown in figure \ref{fig:extretatgbeta46}. As shown, the extrapolation to the required lower values of $m_U^2$ introduces some uncertainty which we quantify later on. Comparing with our treatment of SM-like friction in ref.~\cite{Sopena:2010zz} we find the friction coefficient enhanced by a factor 5 to 10. This seems very plausible, given the couplings and number of degrees of freedom of the light stops compared to e.g. that of the W-bosons.

\begin{table}
\caption{Phenomenological friction coefficient $\eta$ fitted to the 1-loop MSSM wall velocity in ref.~\cite{John:2000zq} for $m_Q = 2000$ GeV and different values of $\tan\beta$, $m_U$ [GeV].}
\centering
\bigskip
\begin{tabular}{c c c c}
\hline\hline
$\tan\beta$ & $m_U^2$ & $v_w$ (John and Schmidt) & Fitted $\eta$\\
\hline
2 & -60$^2$ & 0.060 & 4.58 \\
  & 0     & 0.090 & 3.36 \\
  & +60$^2$ & 0.160 & 1.92 \\
4 & -60$^2$ & 0.080 & 4.35 \\
  & 0     & 0.115 & 3.16 \\
  & +60$^2$ & 0.140 & 2.72 \\
6 & -60$^2$ & 0.085 & 4.65 \\
  & 0     & 0.120 & 3.06 \\
	& +60$^2$ & 0.155 & 2.55 \\
\end{tabular}
\label{table:fitted etas}
\end{table}

\section{Results}

With the extrapolated values of the friction parameter the coupled hydrodynamic equations (\ref{eqn:fluid1})-(\ref{eqn:fluid3}) can be solved using the full 2-loop thermal potential of the LSS. The results for the wall velocity of this calculation are shown in table \ref{table:vw2loop} in two representative cases for $\tan \beta = 4$, $6$ (these values being the most favourable choices as regards obtaining both a conveniently high Higgs mass and a strong phase transition). 

\begin{table}
\caption{Results of wall velocity calculation at 2-loop order for Higgs mass $m_h=112$ GeV, $\tan\beta=4$,$6$ and strength of the phase transition $\frac{v_c}{T_c}$ close to 1. The values of $\xi_{b} = \frac{v_{b}}{T_{b}}$ in the broken symmetry phase are given for comparison.}
\centering
\bigskip
\begin{tabular}{c c c c c}
\hline\hline
$\tan\beta$ & $\xi_c=\frac{v_c}{T_c}$ & $\xi_{b}=\frac{v_{b}}{T_{b}}$ & Fitted $\eta$ & $v_w$\\
\hline
4 & 0.9 & 1.01 & 5.94 & 0.044 \\
  & 1.0 & 1.14 & 6.05 & 0.043 \\
  &     & 1.14 & 4.58 (2-point) & 0.057 \\
  &     & 1.14 & 7.02 (2-point) & 0.037 \\
6 & 0.9 & 1.00 & 6.66 & 0.039 \\
  & 1.0 & 1.14 & 6.79 & 0.038 \\
\end{tabular}
\label{table:vw2loop}
\end{table}

To test our extrapolation of the phenomenological friction parameter we found the highest possible $\eta$ variation found by taking only two calibration points from \cite{John:2000zq} for the case $\tan \beta = 4$. We show this two-point calibration in figure \ref{fig:extretatgbeta46} and give the alternative wall velocities thus found for the case $m_h = 112$ GeV, $\xi_c = \frac{v_c}{T_c} = 1$,  in table \ref{table:vw2loop}. As seen, the wall velocity varies accordingly from 0.037 to 0.057. While this change is noticeable, the prediction of a very subsonic wall velocity in the LSS is robust, and justifies the use of $v_w\sim0.05$ in computations of the resulting baryon asymmetry, e.g.~in ref.~\cite{Konstandin:2005cd}. Keeping our original (3-point) calibration and sampling the most promising areas of parameter space as regards baryogenesis and the Higgs mass we find that the wall velocity varies less than $10 \%$  across the region shown in figure \ref{fig:MSSMparamspace}, staying close to $v_w=0.05$.

\begin{figure}
\begin{center}
\psfrag{xi}{$\xi = v/T$}
\includegraphics{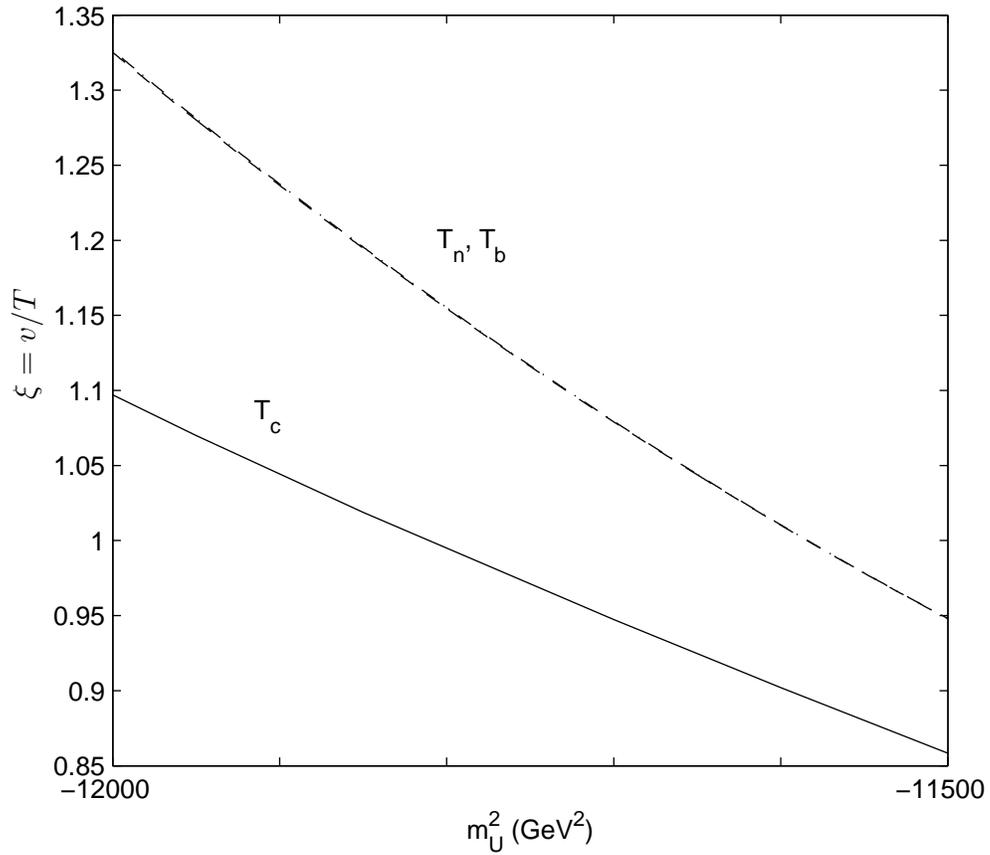}
\caption{Values of $\xi = v/T$ for the case $\tan \beta = 4$, $m_Q = 14,000$ GeV ($m_h \sim 112$ GeV) at the critical and nucleation temperatures and in the broken symmetry phase where sphalerons must be suppressed to avoid washing out the newly-generated baryon asymmetry.}
\label{fig:xiMSSM}
\end{center}
\end{figure}

Relevant for baryon number washout is the value of the Higgs field inside the bubbles, $v_b$, related to the temperature inside the bubbles, $T_b$. The resulting ratio $\xi_b=v_b/T_b$ will be different from the commonly used $\xi_c=v_c/T_c$, obtained from the equilibrium thermal potential at the critical temperature. It is important to understand that the determination of $\xi_b$ requires the solution of the bubble evolution equations in the presence of a plasma, such as eqs.~(\ref{eqn:fluid1})-(\ref{eqn:fluid3}).
We show $\xi$ in figure \ref{fig:xiMSSM} as a function of $m_U^2$ for the case $\tan \beta = 4$, $m_Q = 14,000$. $\xi_b$ is related to $T_b$, while $\xi_c$ is related to $T_c$.
Note that in this case the nucleation temperature, $T_n$, and the temperature inside the bubbles happen to be nearly identical. The value of $\xi_b$ is significantly higher than $\xi_c$, so the commonly used criterion systematically underestimates the strength of the phase transition by at least 10\%.

The low values of the wall velocity found in this setting question some of the assumptions often made when studying the real time history of first-order phase transitions. The bubbles' relatively slow expansion increases the gap between the nucleation temperature and the so-called \textit{finalisation temperature} $T_f$, at which the proportion of space occupied by the growing bubbles reaches unity and the phase transition ends. As an approximation it is usually assumed in the calculation of $T_f$ (see, e.g.~\cite{Anderson:1991zb}) that the bubbles expand at the speed of light. It seems reasonable in our case to quantify the error resulting from that assumption. An alternative choice of e.g.~$v_w = 0.05$ results in a different finalisation temperature $T_f'$. As an extension to our calculation we investigate the variation in the wall velocity introduced by assuming that the temperature of the undisturbed universe outside the bubble is a) the nucleation temperature $T_n$; b) the finalisation temperature calculated in the usual way $T_f$; and c) the finalisation temperature calculated assuming a slow wall velocity $T_f'$. We show the results in table \ref{table:vw2loop} for the case $m_h = 116$ GeV, $m_U^2 = -11900$ GeV$^2$ \footnote{Low values of $m_U^2$ are most favourable for baryogenesis.} with $\tan \beta = 4$, $6$. The choice of finalisation temperature does not introduce a huge variation in $v_w$ but whether we take the temperature of the undisturbed universe as $T_n$ or $T_f$ makes a significant difference. So the bubbles accelerate between the time when they are nucleated and the time when they collide at the end of the phase transition, which is the relevant time for electroweak baryogenesis.
However, it is important to recall that our hydrodynamic treatment assumes bubble expansion at a steady-state velocity and cannot account for bubble wall acceleration, due to the variation in temperature between the beginning and the end of the phase transition or for any other reason.

\begin{table}
\caption{Wall velocity for the case $m_h = 116$ GeV, $m_U^2 = -11900$ GeV$^2$ taking as the temperature of the universe the nucleation temperature $T_n$ and the finalisation temperatures calculated on the assumption $v_w = 1$ ($T_f$), $v_w = 0.05$ ($T_f'$)}
\centering
\bigskip
\begin{tabular}{c c c c c}
\hline\hline
$\tan\beta$ & $T$ & $\xi = \frac{v}{T}$ & Fitted $\eta$ & $v_w$\\
\hline
4 & $T_n$   & 0.92  & 6.10   & 0.044\\
  & $T_f$   & 0.94  &        & 0.051\\
  & $T_f'$ & 0.94  &        & 0.054\\
6 & $T_n$   & 0.95  & 6.88   & 0.039\\
  & $T_f$   & 0.96  &        & 0.045\\
  & $T_f'$ & 0.96  &        & 0.047\\
\end{tabular}
\label{table:vwTnTf}
\end{table}

\section{Conclusions}
We have computed the bubble wall velocity during a strong first-order electroweak phase transition in the light stop scenario of the MSSM. This requires a Higgs mass close to the present lower experimental bound of 115 GeV, a light right-handed stop with a mass around 140 GeV and a heavy left-handed stop with a mass in the multi-TeV range.

 The bubble wall velocity is determined by balancing the pressure inside the expanding bubbles with friction from the hot plasma. We describe the system as a single Higgs field coupled to relativistic hydrodynamics via a phenomenological friction term. The size of this friction term we determine by matching to the wall velocities computed microscopically in ref.~\cite{John:2000zq}. Ref.~\cite{John:2000zq} used the thermal Higgs potential in the 1-loop approximation and was therefore not able to have a strong phase transition at realistic Higgs masses. We use the full 2-loop potential and are able to treat the physically interesting case of $v/T\sim1$ for Higgs masses of around and above 115 GeV. We find a wall velocity $v_w\sim0.05$, not very different from the results obtained in ref.~\cite{John:2000zq} for much weaker phase transitions. The obtained friction coefficient is roughly 5 to 10 times larger than what is found in an SM-like situation \cite{Sopena:2010zz}. It could also be used to treat e.g.~the case of an additional singlet scalar field being added to the light stop scenario. Our result confirms the expectation of slowly moving bubble walls in the light stop scenario, in contrast to ref.~\cite{Megevand:2009gh}, where $v_w\sim0.4$ was found, using a simplified microscopic model of friction.
 
 Having computed the full scalar profile of the expanding bubble, we can more reliably check the criterion for avoiding baryon number washout, by taking the true Higgs field value and temperature inside the expanding bubble, rather than relying on the equilibrium potential at the critical temperature. This way the phase transitions turns out to be about 10\% stronger.
 
The wall velocity is not only crucial for the generated baryon asymmetry, but also for the production of gravitational waves (see, e.g.~\cite{Grojean:2006bp,Huber:2008hg,Huber:2007vva}). However, the gravitational wave signal goes down with the wall velocity, and should therefore not be observable for the MSSM electroweak phase transition.
 
 Finally, we will treat the non-supersymmetric case in more detail in ref.~\cite{HuberSopena}.
 
 \section*{Acknowledgements}
 We like to thank Peter John for clarifications on ref.~\cite{John:2000zq} and Thomas Konstandin for interesting discussions. SH acknowledges support from the Science and Technology Facilities Councel [grant number ST/G000573/1].

\end{document}